# IMPROVING GRAPH-BASED DETECTION OF SINGULAR EVENTS FOR PHOTOCHEMICAL SMOG AGENTS

Rafael Carmona-Cabezas[1, *], Javier Gómez-Gómez[1], Eduardo Gutiérrez de Ravé[1], Elena Sánchez-López[1], João Serrano[2], Francisco José Jiménez-Hornero[1]

[1] Complex Geometry, Patterns and Scaling in Natural and Human Phenomena (GEPENA) Research Group, University of Cordoba, Gregor Mendel Building (3rd floor), Campus Rabanales, 14071 Cordoba, Spain

[2] Mediterranean Institute for Agriculture, Environment and Development (MED), Departamento de Engenharia Rural, Escola de Ciências e Tecnologia, Universidade de Évora, P.O. Box 94, Évora 7002-554, Portugal

* Corresponding author. e-mail: f12carcr@uco.es

ABSTRACT

Recently, a set of graph-based tools have been introduced for the identification of singular events of $O_3$, $NO$ and temperature time series, as well as description of their dynamics. These are based on the use of the Visibility Graphs (VG). In this work, an improvement of the original approach is proposed, being called Upside-Down Visibility Graph (UDVG). It adds the possibility of investigating the singular lowest episodes, instead of the highest. Results confirm the applicability of the new method for describing the multifractal nature of the underlying $O$ , $NO$ , and temperature. Asymmetries in the $NO$ degree distribution are observed, possibly due to the interaction with different chemicals. Furthermore, a comparison of VG and UDVG has been performed and the outcomes show that they describe opposite subsets of the time series (low and high values) as expected. The combination of the results from the two networks is proposed and evaluated, with the aim of obtaining all the information at once. It turns out to be a more complete tool for singularity detection in photochemical time series, which could be a valuable asset for future research.

Abbreviations: VOCs (Volatile Organic Compounds), VG (Visibility Graph), UDVG (Upside-Down Visibility Graph), SP (Shortest Path)

KEYWORDS

- Photochemical smog

- Visibility Graphs

- Singularity detection

1. INTRODUCTION

Among the problems related to atmospheric pollution, there is a matter of special concern studied by environmental scientists in the recent years, the so-called photochemical smog. Also known as "Los Angeles smog", since it was firstly noticed in that city in 1944, as a result of the observed damage on the vegetation (NAPCA, 1970). It can be defined as the accumulation of gases and aerosols as a result of reactions between nitrogen oxides ($NO_x$), certain volatile organic compounds (VOCs) and oxygen under the influence of solar radiation. A wide range of chemicals (ozone, aldehydes or hydrogen peroxides among them) are created in the process (Guicherit and van Dop, 1977). Typically, this phenomenon is more prominent when a city is more populated and warmer. Among the gases involved, there are two which are extensively researched due to the many harms associated to them and their quantitative importance: the tropospheric ozone ($O_3$) and the nitrogen dioxide ($NO$ ), being the second a precursor of the first one. It must be stressed that both of them ($O$  and $NO$ ) have a serious impact on human health (Cheng et al., 2020; Kampa and Castanas, 2008; Yue et al., 2018). Furthermore, a recent study has demonstrated that $O$  produces harsh effects on the economy due to a reduction of the crop yield (Miao et al., 2017).

During the last decades, investigation on complex networks and their applications has been carried out in many works (Boccaletti et al., 2006; Gan et al., 2014; Newman, 2003; Stam, 2010). A complex network can be understood as a graph (a set of nodes and edges as will be further explained) which exhibits nontrivial topological properties and is often used to model and describe real systems. Furthermore, in the recent years there have been a considerable amount of works seeking ways to represent nonlinear time series as complex networks (Zou et al., 2019). This includes manuscripts based on recurrence networks, transition networks and visibility graphs. The main potential of these approaches is the vast number of tools that there exist to analyze networks from a computational perspective. Authors highlight the centrality parameters, since they are essential to this work. They are used to quantify the importance of the nodes within a graph and will be introduced and used later in the text.

Among the new methodologies previously described, there is one that has been recently used to investigate environmental time series (Carmona-Cabezas et al., 2019b; Donner and Donges, 2012; Pierini et al., 2012). This methodology received the denomination of Visibility Graph (VG) algorithm (Lacasa et al., 2008). As it has been demonstrated several times, the complex networks obtained through this method inherit the main features of the original time series and therefore can be used to describe them (Lacasa et al., 2009; Lacasa and Toral, 2010).

Besides describing the nature and main features of the time series, another possibility implies the detection of singularities within these signals. For that purpose, many techniques have been used. One example is the Hölder exponent, which is based on multifractal properties of the system (Loutridis, 2007; Shang et al., 2006). By looking at the information retrieved from the transformed complex network, it is also possible to detect singular points, as it has been explored in several works recently (Bielinskyi and Soloviev, 2018; Carmona-Cabezas et al., 2019b, 2020). In particular, the unusually large values of the cited centrality parameters associated to each node, can provide much of the information that could be derived from the time series.

In the presented work, a new approach is introduced to improve this detection of singular points in a time series from photochemical smog variables (pollutant concentration and temperature), using the VG. The motivation behind it was the fact that regular VG criterion associates the highest connectivity to the points with largest concentration. Therefore, singular events that have low value are overlooked by the original technique. The proposed improvement analyzes the original and inverted series and combines their parameters for a wider point of view.

The pursued aim with this work is to test the application range and possible advantages or pitfalls of the proposed improvement. By doing that, authors intend to explore how this advance could complement the identification of singular episodes of pollutant time series (which could be potentially extended to others apart from $O$ and $NO$ ). Being that the case, future researchers will benefit from a more thorough technique for detecting unusual low and high gas concentrations, with different criteria, as a result.

2. MATERIALS AND METHODS

   2.1. Data

For this work, measurements of tropospheric ozone ($O_3$), nitrogen dioxide ($NO$ ) and temperature have been used. All of them correspond to hourly time series, being recorded in 2017. In the last part of this manuscript, months corresponding to different seasons are selected. The reason for this lies in the fact that, as explained before, this work seeks improving a previous one (Carmona-Cabezas et al., 2020), and therefore the same months have been used for clearer comparison. The station where they were collected is called San Fernando (36°27' N, 6°12' W), which is located in the province of Cádiz (southern Iberian Peninsula) and administered by the Consejería de Medioambiente (Regional Environmental Department) of Andalusia and the European Union.

According to the Köppen-Geiger classification, the zone where the data is collected is labelled as "Csa", as it is most of the Mediterranean basin. "Csa" regions are characterized by warm temperatures with summers that are regularly hot and dry. Furthermore, two of the most important industrial centers in the region (Huelva and Bay of Algeciras) are located relatively close to the study area. As a result of the mentioned conditions, this selected place is propense to accumulation of tropospheric ozone ($O_3$) and nitrogen dioxide ($NO$ ) (Domínguez-López et al., 2014).

## 2.2. Visibility Graph

As it was introduced before, in the last decade, a new method to analyze one dimensional series was introduced (Lacasa et al., 2008). This technique transforms these series into a different mathematical entity: a graph or network. Therefore, it was given the name Visibility Graph, because of its resemblance to the original one used in architecture for space analysis (Lozano-Pérez and Wesley, 1979; Turner et al., 2001). One of the main features of the VG is that it has been demonstrated that it inherits properties of the original time series that it is obtained from (Lacasa et al., 2009, 2008; Lacasa and Toral, 2010). For instance, a periodic series would result on a regular graph after applying it.

In general, a graph can be understood as a set of *nodes* and *edges* that link them. In the context of VG, the nodes correspond to the points in the time series. Thus, it is necessary to stablish the criterion for linking them and so stablishing the *edges*. The basic idea is that two nodes are connected to each other if a line between them can be drawn and it does not pass below any other point in the signal. That is, two points ($t_a$, $y_a$) and ($t_b$, $y_b$) are connected in the graph (have visibility) if any point ($t_c$, $y_c$) between them ($t_a < t_c < t_b$) fulfills:

$$y_c < y_a + (y_b - y_a)\frac{t_c - t_a}{t_b - t_a} \qquad (1)$$

From the VG method described, it is easy to see that the nodes with highest connectivity (also known as *hubs*) will be usually the ones with the unusual greatest values in the time series. This approach comes in handy in order to describe these points with higher magnitude; nevertheless, if one is interested on what happens with the opposite case (i.e. minimal unlikely values), the indicated technique is not suitable for describing them. That is indeed one disadvantage of employing VG for detecting singular points in a time series.

In a recent work, a variation of VG was presented (Soni, 2019) in order to explore new approaches to gain information about a time series. There, the concept of a signed complex network is introduced. The basic idea behind that method is that some of the edges will have a positive sign, while some other will be negative. The regular VG computed as explained before corresponds to the positive edges of this signed graph. On the other hand, the negative connections are obtained also from the regular VG but performed this time over the "upside-down" time series. That is, instead of using the original series $f(t)$, the converted series $-f(t)$ is used. This new graph was employed as a whole, in order to obtain series of clusters from the network and to analyze multivariate correlations, as an extension of previous works (Lacasa et al., 2015; Sannino et al., 2017). Nevertheless, the purpose of the work introduced here is to investigate the possibility of applying this idea for improving the detection of singular points in a time series, such as $O$ and $NO$ concentration, or temperature. For that reason, the positive and negative parts need to be obtained separately, as some of the parameters that will be further explained cannot be retrieved from a signed network (e.g. the betweenness centrality). For clarity reasons, the "positive" network will be given the name of regular VG in the text; while for the "negative" one, the term Upside-Down VG (UDVG) will be used. In Figure 1, an example of the two types of network is shown.

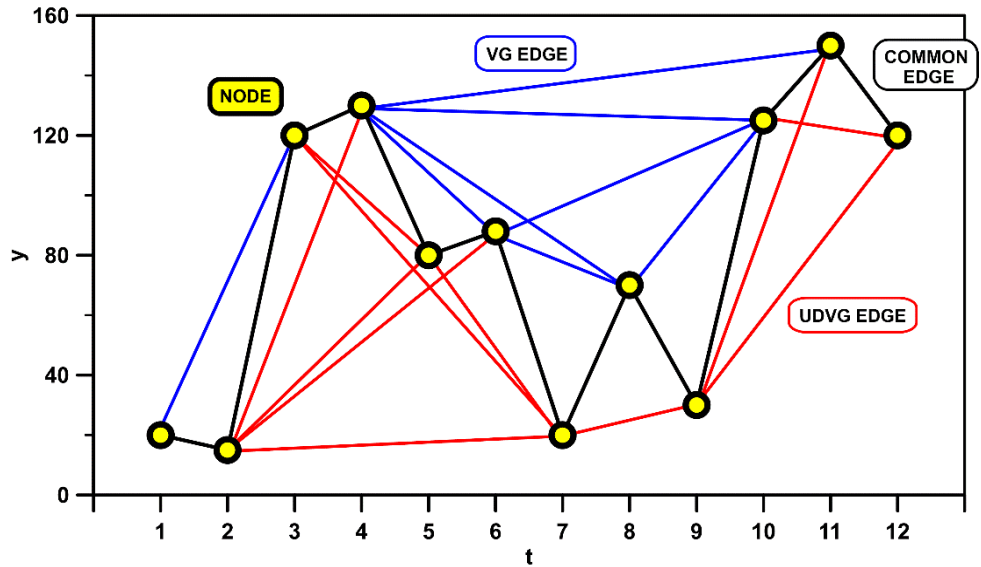

Figure 1: Example of computation of the regular VG (blue lines) and the UDVG (red lines) to a sample time series and resulting graphs. Black lines indicate the common edges.

It must be highlighted as well that all the edges of the two graphs are different, except for those connecting each node to it nearest neighbors in the time series. In Figure 1, the first statement is seen by looking at the blue and red edges, while the second one reflects in the black ones that both regular VG and UDVG have in common. In other words, the elements of the adjacency matrices fulfill: $a_{ij}^{VG} + a_{ij}^{UDVG} \leq 1; \forall j \neq i \pm 1$. This means that the elements surrounding the main diagonal are equal and the others cannot be $a_{ij} = 1$ simultaneously in the two matrices.

## 2.3. Centrality parameters

One of the most widely used approaches to characterize graphs and complex networks is based on the analysis of the most important nodes within. It is done by employing the so-called centrality parameters, which are evaluated at each node, giving an idea about how "central" each one is, in relation to the rest of them. This concept was firstly used for studying social networks and transferred to other fields of research afterwards (Agryzkov et al., 2019; Joyce et al., 2010; Liu et al., 2015). The actual meaning of a central node may vary depending on the actual parameter used to evaluate the network. Here, authors focus on three of them: the degree, betweenness and closeness centrality, which have been used to describe physical systems in previous works (Carmona-Cabezas et al., 2020; Donner and Donges, 2012; Mali et al., 2018).

The first centrality parameter that will be explained is the degree. In a graph, the number of edges which are connected to a given node $i$ is defined as the degree of that node ($k_i$), i.e., $k_i = \sum_j a_{ij}$, being $a_{ij}$ the elements of the adjacency matrix. Once the degree for each node is obtained, the degree distribution $P(k)$ can be computed. This quantity has been proven to be able to characterize the nature of the studied signal (Lacasa et al., 2008; Mali et al., 2018). In fact, degree distributions that can be adjusted to a power law $P(k) \propto k^{-\gamma}$ correspond to scale free networks which comes from fractal series, as it was discussed by (Lacasa et al., 2009; Lacasa and Toral, 2010). The reason for this is the effect of hub repulsion (Song et al., 2006). A hub is a node from a graph with unlikely greater number of links, and so, higher degree. Therefore, the right tail of degree distributions is dominated by these nodes and, after being represented in a log-log plot, they can be fitted by a simple linear regression.

The other two employed parameters cannot be understood without defining the shortest path (SP) quantity first. SP is a measurement of the number of different edges that connect two distant nodes. Given a pair of nodes $(i, j)$, different possible paths between them are available. Some of them (not necessarily unique) will have the minimum possible number of edges and, thus, they will be the minimal paths known as SP. It must be regarded that it has an important presence in the definition of the betweenness and closeness centrality. The betweenness of a node $i$ can be computed by the following expression:

$$b_i = \sum_{\substack{j=1 \\ j \neq i}}^{N} \sum_{\substack{k=1 \\ k \neq i,j}}^{N} \frac{n_{jk}(i)}{n_{jk}} \qquad (2)$$

Where $n_{jk}$ is the number of SP's from node $j$ to $k$, whereas $n_{jk}(i)$ is the number of those SP's that contain the node $i$. A high betweenness can be interpreted as a node which is passed through by SP's connecting the rest of nodes.

Lastly, the closeness centrality is obtained as shown in the following expression:

$$c_i = \frac{1}{\sum_{j=1}^{N} d_{i,j}} \qquad (3)$$

There, the closeness of each node $c_i$ is computed from the so-called distance matrix $D$, where each element $d_{i,j}$ corresponds to the SP from node $i$ to $j$. Therefore, this quantity accounts for how close a given node is to the rest of the network, in terms of edges needed for other nodes to be reached.

3. RESULTS AND DISCUSSION

    3.1. Degree distributions

After the proposed methodology has been explained, authors have analyzed firstly how the UDVG degree distribution differs from the regular VG with the same time series. It served as a preliminary study, before tackling the identification of relevant points in the signal, which is the main objective of this work.

Figure 2 represents two theoretical time series have been employed to test the method. The first one of them is obtained from a fractional Brownian motion with Hurst exponent $H = 0.5$ and $10^4$ points. The second one corresponds to a random series with $10^5$ points. The reason for choosing them is that they are standard well-known series that are frequently used within this type of studies with VGs. This figure shows the series (a and b) and their respective degree distribution computed for both approaches (c and d). It can be inferred that the distributions that arise from using the UDVG are almost identical to the VG ones. Therefore, at least for these types of series, the VG and UDVG degree distributions describe the same properties of the underlying time series.

In the case of the fractional Brownian motion, they also present curves which can be adjusted to the same power law. Thus, this might indicate that UDVG would be also suitable for describing scale-free networks, such as those extracted from these type of series, which are fractal (Lacasa et al., 2009). For the random time series, the result is a distribution with a tail that follows an exponential trend, as expected (Lacasa et al., 2008).

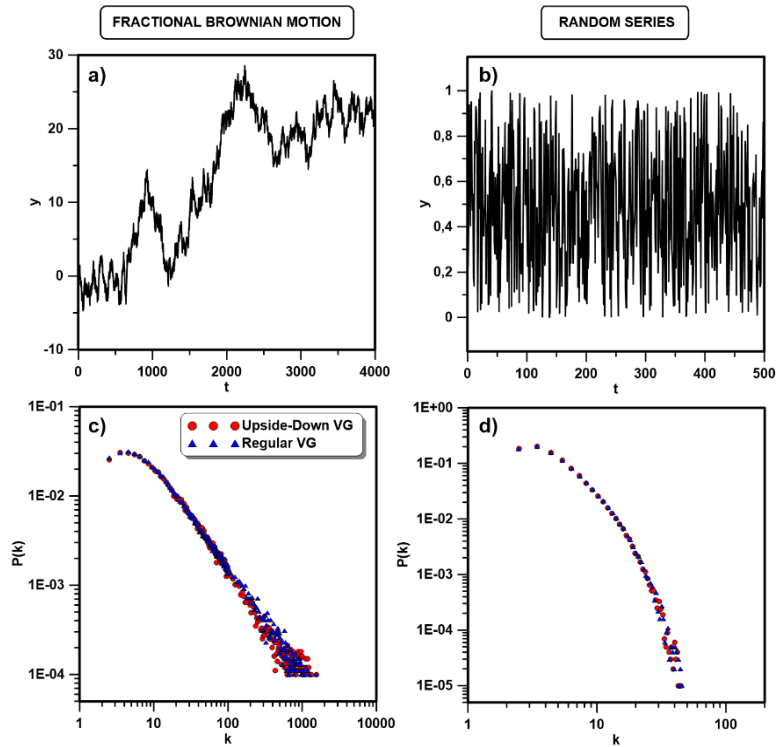

Figure 2: Top: fractional Brownian motion signal with Hurst exponent $H = 0.5$ and $10^4$ points (a) and 500 points from a random series (b). Bottom: The degree distribution computed from the complex networks obtained for both series, by employing the VG and UDVG.

Once it was observed that UDVG and VG obtain the same results for the theoretical time series, authors have tested the photochemical time series, which are the focus of this study. These correspond to three different signals: two from $O$ and $NO_2$ concentration, and the other one temperature, all of them from the same year (2017), as previously stated. The reason behind choosing one complete year in this particular part of the study is that for a reliable comparison of degree distributions, a considerable amount of points in the time series is required (Carmona-Cabezas et al., 2019a). Since the resolution of the measurements is one hour, it has been observed after several tests that taking only monthly samples for comparison could give misleading results. It should be underscored that the same is not true for the later analysis of singular episodes, since that is a local study and does not depend in the extension of the pollutant time series.

The three signals are depicted in the upper part of Figure 3 (a1, b1 and c1). Conversely to what was observed in the previous case, now Figure 3 (a2, b2 and c2) display slight differences between the degree distributions of the classic VG and the UDVG. Nevertheless, a clear power law behavior is observed in every case. This is in accordance with previous works, where ozone ($O_3$) and nitrogen dioxide ($NO$) time series exhibited scale-free behavior, as a consequence of the multifractal nature of its dynamics (He, 2017; Pavón-Domínguez et al., 2015). The observed contrasts are more pronounced in the case of the nitrogen dioxide ($NO$) concentration time series, clearly showing a marked difference in the slope of the distribution tail (the $\gamma$-exponent). Authors attribute this effect to the difference between the three underlying time series. The ground-level ozone signal exhibits a pattern that equally presents singular minima and maxima, and the same can be said about the temperature. Therefore, the frequency distributions of their concentrations will have roughly symmetric shapes. On the other hand, the same cannot be argued for the nitrogen dioxide ($NO$) concentration. Minima and maxima values are not distributed evenly along the time series, which is clear in Figure 3c. The maxima are rather infrequent and singular in comparison to the minima, which are much more common, as most of the values are very close to zero. Therefore, one could expect the probability distribution of the concentration of nitrogen dioxide ($NO$) to be non-symmetric. To investigate that, the most suitable method is to inspect the skewness ($S$) for each time series. This quantity describes the asymmetry of the probability distribution of a given real measure around its mean. When skewness is equal to zero, it means that the distribution is symmetric respect to its mean, being the opposite case ($S \neq 0$) for non-symmetric distributions. In Figure 3, probability distributions of concentrations and temperature are depicted with their respective skewness value. In the case of $O$ and temperature (Figure 3 a3 and c3), the distributions are almost symmetric, as mentioned before, with skewness close to zero ($S_{O_3} = -0.28$ and $S_{temp} = 0.21$). Despite this, a mild deviation between the regular and inverted distributions can be observed, leading to the low negative skewness that is observed. On the contrary, a positive skewness value ($S_{NO_2} = 2.32$) of nitrogen dioxide ($NO$) concentration is clearly seen (Figure 3b3), i.e. low values with respect to the mean are highly frequent. Therefore, the concentration of nitrogen dioxide ($NO$) in San Fernando reaches low peaks many times during the month, while the high accumulations of this noxious gas are much rarer.

For the case of temperature, this symmetry can be interpreted as the relatively regular behavior of day and night values, meaning that the appearances of singular episodes of low and high temperature will be linked during the year, depending on the meteorological conditions of each season. On the other hand, one could expect this difference between nitrogen dioxide ($NO$) and tropospheric ozone ($O_3$), regarding the symmetry of the degree distribution. Both gases are correlated through the simplified photochemical reaction $NO + O \leftrightarrow O + NO$. Production and destruction of ozone will occur during day and night times respectively. The photolysis that leads to the ozone accumulation and the reach of the photostationary state regularly happens during mid-day, when there is radiation available. The sense of the reaction is reverted during nighttime in the absence of light. Although the quantitative concentration levels may vary depending on factors such as wind speed, temperature or mixing height, the distributions of maxima and minima could be expected to be symmetric as it is seen for the $O_3$. However, the nitrogen dioxide ($NO$) intervenes in other reactions that could lead to the appearance of singular minima in its concentration. One example is the aldehyde production through interaction with VOCs, which results on a lower rate of $NO$-$NO$ reaction. For a deeper understanding of this, further analysis with $NO$ and VOCs time series would be necessary.

Authors would like to point out that, for the previous theoretical series this relation is also observed, being their computed skewness values very close to zero ($S_{brownian} = 0.13$ and $S_{random} = 3.10 \cdot 10^{-4}$), as expected since their distribution where almost perfectly coincident for VG and UDVG.

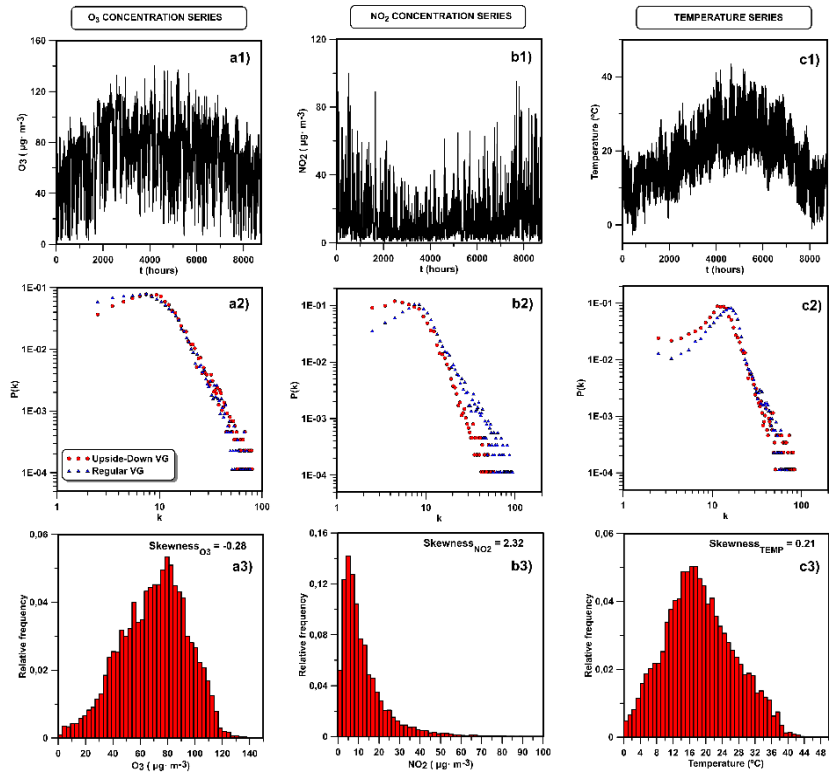

Figure 3: Top: Ozone ($O_3$), nitrogen dioxide ($NO_2$) concentration and temperature annual temporal series (a1, b1 and c1). Middle: Degree distributions computed with regular VG and UDVG (a2, b2 and c2). Bottom: frequency distributions of the pollutant concentrations and temperature time series (a3, b3 and c3).

### 3.2. Identification of hubs

After the preliminary study of probability distributions has been carried out, a pointwise study of $NO$ and $O$ concentrations and temperature is undertaken here. Figure 4 depicts a comparison between the hubs computed by applying the VG on the unvaried time series and those of the inverted one. Now only one real time series is shown, because the actual interest here is to observe the differences between UDVG and regular VG when detecting the singular extremes. In this case, only one month (July) from the ozone concentration time series was chosen for the sake of clarity (see Figure 4a). This month was chosen because, in this location, July is the period of the year were the most severe episodes of ozone pollution occur. In the next two figures (Figure 4b and c), the normalized betweenness and degree values are shown for both networks (blue is for the original VG, while red for the UDVG). Only these two centrality parameters were chosen in this case because they have clearer signals. The three centrality parameters presented in the methodology section of this work will be used in later discussions.

It can be regarded in Figure 4 the fact that both networks (the regular and inverted one) are able to identify extrema in the time series in a complementary manner, as anticipated. While the regular VG hubs correspond maximal episodes of tropospheric ozone concentration (which has been already used), the UDVG obtained ones do the same with minima of the concentration. These latter correspond to the nighttime, when the photochemical reaction is unbalanced towards $NO$ formation in the absence of radiation. The actual physical interpretation of the different centrality parameters can be observed in the previous related work (Carmona-Cabezas et al., 2020) for the regular VG hubs. Additionally, it will be explained for the UDVG case in the last figures.

Moreover, it must be noticed how the hubs from betweenness coincide with those of the degree, while the opposite case is not always true. Therefore, the first one may be a more selective approach to identify singular nodes in a signal, as it has been discussed in a previous work (Carmona-Cabezas et al., 2019b). This filtering feature might be useful for the use of this technique on environmental series where the density of extrema is considerably high.

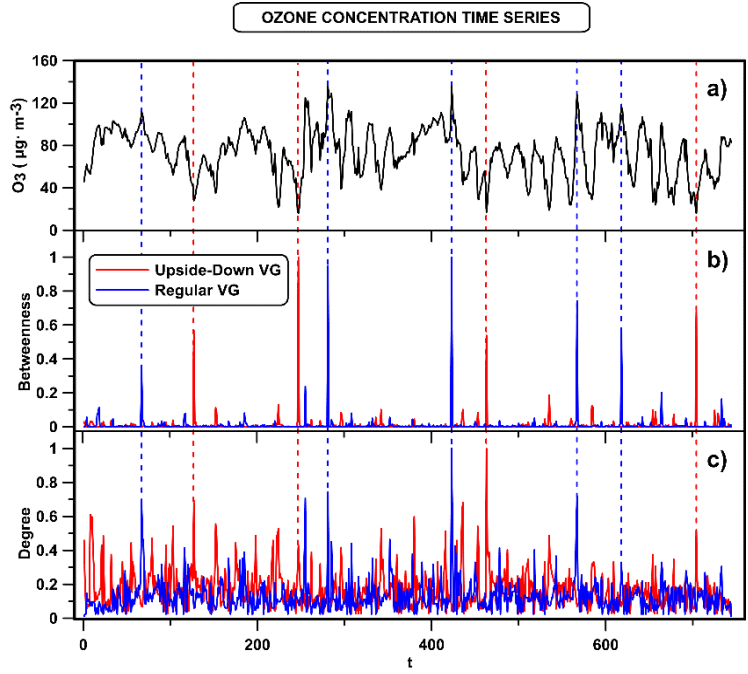

Figure 4: a) Ozone concentration time series from a selected month (July). b) and c) Normalized betweenness and degree centrality values, respectively, for each point in the time series, from the two graphs studied (VG in blue and UDVG in red). The dashed lines are used to highlight the hubs positions and compare them in the three plots.

Once the difference between the VG and UDVG hubs has been discussed, authors propose an approach for combining the information given by both networks. The aim is to yield a more complete technique for future investigations to analyze pollutant time series. The combined parameters tested here simply consist on adding each VG centrality parameter to the opposite of the one computed using UDVG. To make it clearer, for the betweenness, degree and closeness:

$$\begin{cases} b_i{}^{comb} = b_i^{VG} - b_i^{UDVG} \\ k_i{}^{comb} = k_i^{VG} - k_i^{UDVG} \\ c_i{}^{comb} = c_i^{VG} - c_i^{UDVG} \end{cases} \quad (4)$$

This transformation is useful for the identification of singularities or extreme values, considering both the minima and maxima values. It is based on the fact that when the VG maps a hub, the UDVG will not, since they are complementary, as exposed in the Methodology section. Thus, the hubs information is not lost by this procedure, because their values will not be canceled out for the case of extremes. Consequently, it improves the differentiation between regular and singular values. This results in the derived combined degree signal being smoother and clearer than in the separated case. For the combined betweenness, there will be almost no difference in the smoothness, since the values that do not correspond to skyline hubs are practically zero in any case.

For clarity reasons, the same structure as in a previous work (Carmona-Cabezas et al., 2020) has been followed for the plots. Hence, the combined betweenness is computed first and from it, the five most pronounced peaks are chosen automatically. Equivalent results can be yielded by selecting a greater number of peaks, as it has been tested. A criterion for it was indicated in the mentioned previous work (Carmona-Cabezas et al., 2020). Afterwards, the remaining centrality measures were analyzed in the positions where the first peaks are located. It must be highlighted that all the plotted parameters are normalized to the maximum absolute value of each one, for the sake of comparison.

In Figure 5, this explained procedure is performed using the series of tropospheric ozone previously introduced. As it is easily seen, the accordance between the different studied parameters is adequate, as it was expected. The combination of the VG and UDVG still preserve the capability to identify extrema by the different centrality parameters. The smoothest series corresponds to that of the betweenness as previously explained, followed by the degree and finally by the closeness. It is in accordance to what was observed using the less complete method in the previous paper (Carmona-Cabezas et al., 2020). It must be stressed that the order of the magnitude of the different peaks is not conserved in the different combined centrality parameters. For instance, in Figure 5c, the peak 2 is the most negative one, while in Figure 5d and Figure 5e are the peaks 3 and 5 respectively. This is due to the different physical meanings of each parameter related to the concentration time series. Therefore, this should be taken in consideration if different parameters are used to compare ozone (or other pollutant) extreme concentration episodes in future studies.

The first one of them is the combined betweenness (Figure *5*c). In order to understand the usefulness of this parameter to the photochemical pollution, it must be pointed out that in a previous study (Carmona-Cabezas et al., 2019b) skyline hubs were related to values of the series which can give more information about its upper envelope. In short, one of the detected $O_3$ singularities may be considered as an unlikely high episode of ozone concentration in relation to other maxima in the same series. This means that if ozone daily maximal concentrations were raising for several consecutive days, a peak in the betweenness indicates that after this encountered, the trend is likely to change to a downwards one. Conversely, translating this to the inverted series (and the resulting UDVG), the same could be inferred about minimal night $O_3$ concentrations. Environmentally speaking, a change in the tendency could be a pointer to an alteration of the previous ambient conditions that would lead to an abnormal shift in the height of the mixing layer, for instance. Therefore, the combined betweenness can serve as a more complete warning tool, pointing changes in the conditions that affect the temporal evolution of pollutants concentration, while the previous approach would only yield insight on the upper one, limiting the analysis.

Regarding the next complex network indicator, the combined degree (Figure 5d), many works have been devoted to its study (Pierini et al., 2012; Zhou et al., 2017). It is known that a degree hub is associated to a specially high ozone concentration episode (Carmona-Cabezas et al., 2019a). At the position where the hubs are encountered, the gas has reached a peculiarly high concentration. This condition is less restrictive, as every day it is fulfilled. As a result, the number of this type of peaks is greater, compared to the betweenness. In this case, a peak would not be necessarily associated to a change in the prior tendency of the $O$ concentrations. When the regular VG results are combined with the UDVG, the unlikely low values of concentration can be identified as well. Again, the identification of rare concentrations of $O$ is improved by this combination, getting at the same time the information from low and high values from one single parameter. Here, the sense of "singularity" in the ozone is referred only to its magnitude, and not to the trend of the previous and posterior days, as in the betweenness.

Figure 5e) illustrates the closeness centrality results. In previous research, this quantity was mainly used for theoretical purposes. Nonetheless, it was demonstrated recently that it could identify singularities as the previous ones, but with a different criterion (Carmona-Cabezas et al., 2020). The peaks of this magnitude were related to high concentrations of ozone episodes surrounded by concave up tendency. This quantity was found to be noisier than betweenness and degree, and so it is as well here. As in the previous parameters, now the combined quantity (more specifically the negative part), gives additionally information about the points where a minimal rare concentration value is found, surrounded by a concave down accumulation of values (the reversed shape with respect to the regular VG). In the context of photochemical pollution, it would mean that this could be used to identify daily high concentrations (during the photostationary state) that somehow drop, due for instance to unexpected atmospheric conditions.

The selected minima correspond to the 6$^{th}$, 11$^{th}$, 20$^{th}$, 23$^{rd}$ and 30$^{th}$ of July, all of them occurring between 6:00 AM and 7:00 AM (UTC+1) as could be expected. The photochemical reaction is reverted during the nighttime and most of the tropospheric ozone ($O_3$) (produced during the previous day) is recombined with $NO$ to yield $NO$ in the absence of light. After this time, there is radiation available and its concentration has an upward trend. In the previous work, the singular high episodes between 2:00 PM and 6:00 PM (UTC+1), which corresponds to the opposite case.

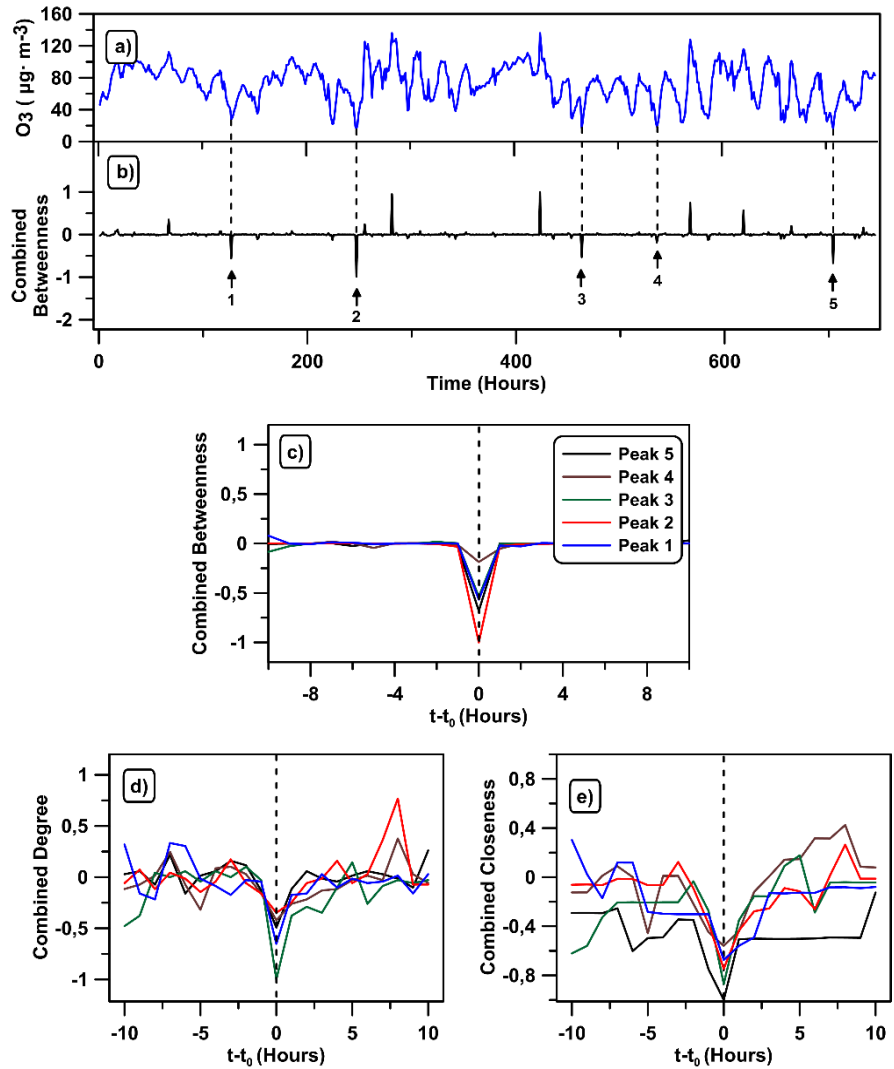

Figure 5: Ozone concentration time series (a) with the combined betweenness computed from the UDVG and VG (b). Plots from c) to e) show the complex network indicators: betweenness, degree and closeness in the selected five negative peaks.

The next graph (Figure 6) shows the results obtained for the nitrogen dioxide ($NO_2$) concentration time series, which as seen before, has a different minima and maxima behavior. For this study, the studied month is January as in the previous work, since in this region that is period of the year when less photochemical activity takes place. Therefore, reactions with other chemicals (such as VOCs to yield aldehydes) could play a more important role, leading to more singular extrema.

Once again, there is a good fit between the different parameters, although in this case, the combined degree is noisier and not as clear as before. This might be caused by the accumulation of low values of nitrogen dioxide ($NO$) concentration that make the distribution to be more asymmetric, as discussed (see Figure *3*). The greater number of reactions that involve $NO$ might increase the number of singularities, being the degree noisier as a result.

In this case, the selected concentrations of $NO$ are in the $7^{th}$, $12^{th}$, $18^{th}$, $27^{th}$ and $31^{st}$ of January, between 2:00 AM and 5:00 AM (UTC+1). Regarding the high singularities investigated in the previous work, there was no consistent time frame where it could be encountered. Also, it is well known that there is a marked difference between concentrations during weekends and weekdays (Qin, 2004), which could be another possible cause for this uncertainty.

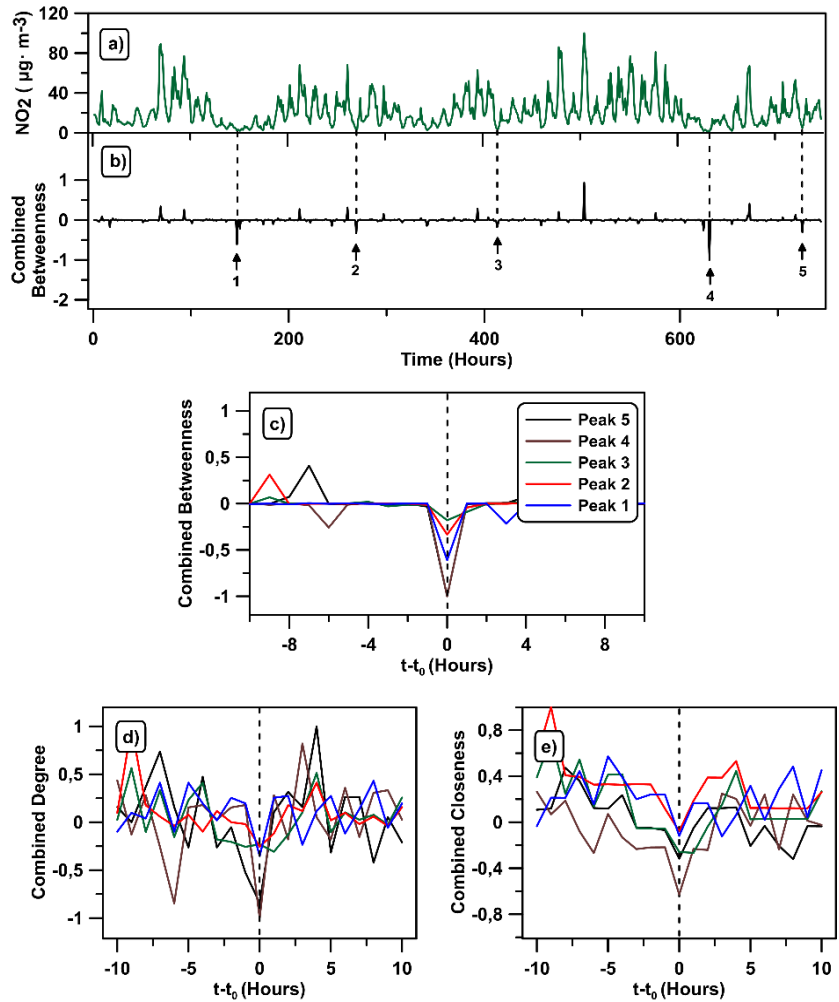

Figure 6: Nitrogen dioxide $(NO_2)$ concentration time series (a) with the combined betweenness computed from the UDVG and VG (b). Plots from c) to e) show the complex network indicators: betweenness, degree and closeness in the selected five negative peaks.

Finally, in Figure 7 the temperature time series is studied locally as in the previous two cases. Now the selected month is October, in order to observe singular episodes of this quantity. Due to the oceanic influence, the temperature is stable throughout almost all the year. Nevertheless, it is more unstable in autumn in this area, as discussed in previous works (Dueñas et al., 2004).

It is clearly seen that for temperature there is also concordance between the combined betweenness and the rest of combined centrality parameters. Now the combined degree signal has less noise than in the case of $NO$, except for Peak 2. This one is more difficult to identify due to the fact that there are two betweenness peaks very close to each other (see Figure 7 b).

The temperature singular minima that have been selected, following the previous criterion, correspond to the $6^{th}$, $11^{th}$, $15^{th}$, $25^{th}$ and $29^{th}$ of October, between 5:00 AM and 8:00 AM (UTC+1). It could be easily expected, since it is the time when the minimum temperature is reached every day. Even during the days in which the temperature becomes more unpredictable (around the middle part of the month), these minima can be observed with a relatively constant frequency.

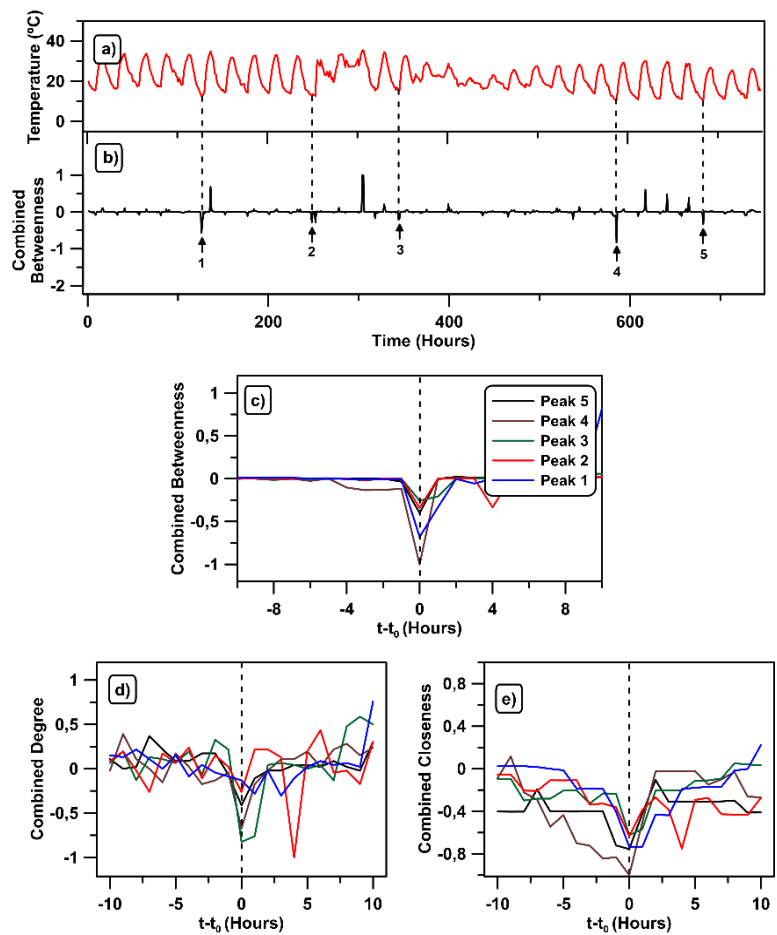

Figure 7: Temperature time series (a) with the combined betweenness computed from the UDVG and VG (b). Plots from c) to e) show the complex network indicators: betweenness, degree and closeness in the selected five negative peaks.

## 4. CONCLUSIONS

An improvement of a singularity detection technique is tested for its application on photochemical time series in this manuscript. It adds the possibility of describing singular minima and maximal singular values at the same time, making it a more complete tool. Authors believe that it may have a great potential for monitoring and analyzing pollutant and atmospheric time series in the future.

The degree distributions obtained have been compared, proving that UDVG inherits the nature of the original $NO_2$, $O$ and temperature time series. Moreover, different theoretical series have been tested, proving the suitability of both VG and UDVG. It has been found that those distribution are coincident for tropospheric ozone ($O_3$) and temperature, while they are not for the nitrogen dioxide ($NO_2$). Their disparity has been related to the greater number of reactions that involve $NO_2$, such us its interaction with VOCs to yield aldehydes. This must be investigated more in detail in a future study, applying different complex networks tools developed to series of $NO_x$, VOCs and $O$ at the same time.

Furthermore, the usefulness of UDVG for singular minima detection has been successfully proven on the $NO_2$, $O$ and temperature series. The combination of VG and UDVG parameters (degree, betweenness and closeness) is proposed as a more exhaustive method, compared to only employing VG. Due to their complementary nature, these combinations store the original information of the most central nodes, showing all the relevant information at a glance. To authors' mind, this can widen the range of the research applications of complex networks for photochemical pollution in a future.

## 5. ACKNOWLEDGEMENTS

The FLAE approach for the sequence of authors is applied in this work. Authors gratefully acknowledge the support of the Andalusian Research Plan Group TEP-957 and the XXIII research program (2018) of the University of Cordoba. Rafael Carmona-Cabezas thanks the Mediterranean Institute for Agriculture, Environment and Development (MED) for its collaboration.

# CREDIT AUTHOR STATEMENT

**Rafael Carmona-Cabezas:** Conceptualization, Methodology, Software, Validation, Formal Analysis, Data Curation, Investigation, Original Draft, Review and editing

**Javier Gómez-Gómez:** Software, Investigation, Original Draft

**Eduardo Gutiérrez de Ravé:** Project Administration, Funding Acquisition, Supervision, Review and editing

**Elena Sánchez-López:** Visualization

**João Serrano:** Resources

**Francisco José Jiménez-Hornero:** Project Administration, Funding Acquisition, Supervision, Review and editing

**Declaration of interests**

☒ The authors declare that they have no known competing financial interests or personal relationships that could have appeared to influence the work reported in this paper.

☐ The authors declare the following financial interests/personal relationships which may be considered as potential competing interests:

# **HIGHLIGHTS**

- Detection of singularities using graphs is improved by taking the inverted series.
- Maxima and minima of pollutant series are identified by VG and UDVG respectively.
- Asymmetries in the distribution of $NO$ might be caused by reaction with VOCs.
- $NO$ singularity identification is more difficult due to its more complex dynamics.
- A more complete analysis tool is obtained by combining both approaches.

**Graphical Abstract**

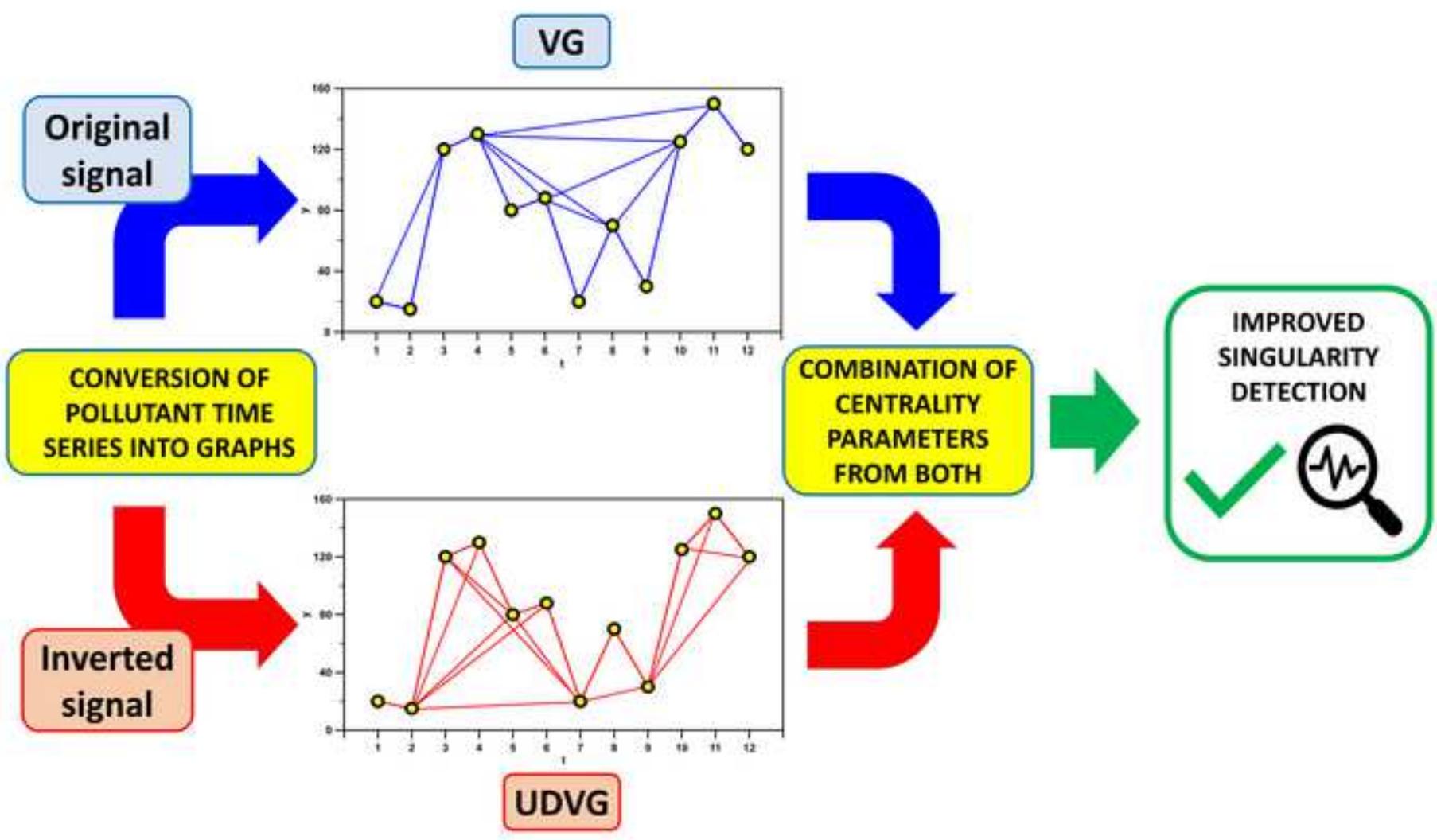